\documentclass[final,3p]{elsarticle}

\usepackage[]{natbib}
\biboptions{sort&compress}
\usepackage{amsmath}
\usepackage{amssymb}
\usepackage{url}
\usepackage{amsthm}
\usepackage{mathtools}
\usepackage{xcolor}
\usepackage{subfig}

\usepackage[colorinlistoftodos,prependcaption,textsize=tiny]{todonotes}
\usepackage[linesnumbered, ruled]{algorithm2e}

\usepackage{float}

\begin{document}

\begin{frontmatter}
\title{Mixture Components Inference for Sparse Regression: Introduction and Application for Estimation of Neuronal Signal from fMRI BOLD}

\author[1,2,3]{Anna Pidnebesna}
\ead{pidnebesna@cs.cas.cz}
\author[2]{Iveta Fajnerov\'{a}}
\ead{iveta.fajnerova@nudz.cz}
\author[2]{Ji\v{r}\'{i} Hor\'{a}\v{c}ek} 	
\ead{jiri.horacek@nudz.cz}
\author[1,2]{Jaroslav Hlinka\corref{cor1}}
\ead{hlinka@cs.cas.cz}

\address[1]{ Institute of Computer Science of the Czech Academy of Sciences, Pod Vod\'{a}renskou v\v{e}\v{z}\'{i} 271/2, 182 07 Prague, Czech Republic}
\address[2]{ National Institute of Mental Health, Topolov\'{a} 748, 250 67 Klecany, Czech Republic}
\address[3]{ Faculty of Electrical Engineering, Czech Technical University, Technick\'{a} 1902/2, 166 27 Prague, Czech Republic}

\cortext[cor1]{Corresponding author}

\vspace{-1em}

  \date{\today}


\begin{abstract}
Sparse linear regression methods including the well-known LASSO and the Dantzig selector have become ubiquitous in the engineering practice, including in medical imaging. Among other tasks, they have been successfully applied for the estimation of neuronal activity from functional magnetic resonance data without prior knowledge of the stimulus or activation timing, utilizing an approximate knowledge of the hemodynamic response to local neuronal activity. These methods work by generating a parametric family of solutions with different sparsity, among which an ultimate choice is made using an information criteria. 
We propose a novel approach, that instead of selecting a single option from the family of regularized solutions, utilizes the whole family of such sparse regression solutions. Namely, their ensemble provides a first approximation of probability of activation at each time-point, and together with the conditional neuronal activity distributions estimated with the theory of mixtures with varying concentrations, they serve as the inputs to a Bayes classifier eventually deciding on the verity of activation at each time-point. 
We show in extensive numerical simulations that this new method performs favourably in comparison with standard approaches in a range of realistic scenarios. This is mainly due to the avoidance of overfitting and underfitting that commonly plague the solutions based on sparse regression combined with model selection methods, including the corrected Akaike Information Criterion. This advantage is finally documented in selected fMRI task datasets. 
\end{abstract}%


\end{frontmatter}

\section{Introduction}

The most common approach to fMRI measurement is using the Blood Oxygenation Level Dependent (BOLD) contrast, see~\cite{Ogawa1990}.
Soon after its discovery, it was shown that it reliably reflects changes in local brain activity related to spontaneous or stimulation-triggered cognitive processing~\cite{Ogawa1992}.
However, the BOLD signal is only indirectly related to the underlying neuronal activity. As its name suggests, it reflects the changes in the blood oxygenation level; these happen after the triggering neuronal activity and constitute a considerably delayed and smoothed response that can be in the first approximation modelled as the output of a linear filter represented by a hemodynamic response function. Combined with a range of noise sources in fMRI~\cite{Bianciardi2009}, statistical inference of neuronal activity from fMRI BOLD signal poses a serious signal analysis challenge.

The common fMRI analysis approach is to impose strong prior  on the timing of the tentative neuronal activity, assuming that it amounts to a predefined time-course reflecting the presence of particular controlled experimental manipulation. This expected activation time course is convolved with a template hemodynamic response and used as a predictor in a linear regression model of the observed BOLD fMRI signal. 

Yet there are still many scenarios when this approach is not suitable, typically when the precise timing of the presumed neuronal activity is not known as it is not perfectly locked to a predefined task paradigm, but rather inter- and intra-individually variable, self-paced or ultimately spontaneously occurring. The importance of having tools  available to analyse such less constrained fMRI BOLD data (compared to regular block and event-related designs) is increasingly recognized as a valuable complement of the standard analysis methods~\cite{Gaudes2013}.


This approach is opposite to the common analysis -- instead of convolving the expected neuronal activity with the hemodynamic response to obtain an approximate expected BOLD fMRI signal (and comparing it with the observed BOLD fMRI via linear regression), the observed signal is \emph{deconvolved} to obtain an estimate of the underlying neuronal signal. 
This 'deconvolution' task can be formulated as a linear regression problem, solvable by various approaches; among them the so-called sparse regression, that attempts to preclude overfitting of noise by inducing penalties for marking too many activations in the data, is most common. The most used sparse regression methods are the LASSO (least absolute shrinkage and selection operator), see~\cite{Tibshirani1996}, and the Dantzig Selector, 
proposed in~\cite{Candes2007}. 
These methods output a parametric family of fitted models (parameterized by a free parameter determining the strength of the regularization and therefore the sparsity of the obtained solution). Thus a suitable model selection procedure must be used, typically the Akaike Information Criteria (AIC)~\cite{Akaike1974} or Bayes Information Criteria (BIC)~\cite{Schwarz1978}. 

While these model selection criteria have been extensively and successfully applied in many fields, they have their drawbacks. In particular, for small sample size AIC tends to overfit the data, giving a too high model order, whereas the BIC-selected model may underfit data, see~\cite{Burnham2004}. Therefore the use of the adjusted Akaike Information Criterion (AICc)~\cite{Sugiura1978,Hurvich1989} was recommended instead of AIC. 

The use of sparse linear regression (specifically the Dantzig Selector) for neuronal signal estimation is described in~\cite{Gaudes2013}.
The authors compare AIC and BIC for the neuronal signal estimation in a simulation study and show that AIC fails to estimate the sparse stimulus signal. Therefore they suggest using the BIC. However, the simulated stimulus was very sparse in the study, suggesting that in the case of  less sparse activations, BIC might be underfitting the data.

We assessed the performance of fMRI deconvolution methods including the canonical LASSO and the Dantzig Selector regularization combined with the most common criteria (AIC, BIC and AICc), arriving at a conclusion that their performance is far from perfect under realistic scenarios, possibly due to nonrobustness of the model selection procedure. 
In this work we thus propose an alternative approach to the estimation, based on the Bayes classifier and theory of mixtures with varying concentrations, that provides more accurate and robust results both in simulations and on a real data sample.
The described algorithm has been implemented into an updated version of software BRAD~\cite{Pidnebesna2018a}.
The work is organized as follows. The model of BOLD signal,  is described in Section~\ref{sec:data}. The practical motivation of the study and theory overview is given in Section~\ref{sec:theory}. In Section~\ref{sec:mic} the new approach is proposed, and methodology and simulation study are described. Finally, the real data application is found in Section~\ref{sec:experiments}.
	
\section{Data Model}
\label{sec:data}

We model the measured BOLD signal $y$ as a result of a hemodynamic response to local brain activity $s$. The neuronal activity is represented by (a sparse) vector of activations; we assume there is no a priori information about stimulus timing. 
In line with the common fMRI BOLD modelling approaches, we assume that the measured signal is a result of the convolution of the neuronal activity and the hemodynamic response function (HRF), in mathematical notation:
\begin{equation}
\label{eq:conv}
y = h \star s + e, \quad y_k = \sum_{i=0}^{N_h-1}h_is_{k-i} + e_k,
\quad k \in \{1,..,N\},
\end{equation} 
where $k \in \{1,..,N\}$ indexes the time, 
$\star$ denotes the convolution, $y,s,e$ are $N\times 1$ vectors denoting the measured signal, the underlying brain activity signal and the observational noise, respectively. The vector $h$ denotes the HRF, where the length $N_h$ is smaller than the length of the measured signal: $N_h<N$. We assume that $e$ is white noise, independent and identically distributed in time. 
The model can be rewritten as
\begin{equation}
\label{eq:regr}
y = Hs+e, 
\end{equation}
where 
$H$ is the Toeplitz convolution matrix of size $N \times N$ corresponding to the HRF $h$ (see~\cite{Gray2006}).

\section{Motivation/Theoretical background}
\label{sec:theory}

	\subsection{Sparse Regression Methods}

We use the two most prominent sparse regression methods - the LASSO (least absolute shrinkage and selection operator) and the Dantzig Selector as the benchmarks. 
LASSO estimator was proposed in~\cite{Tibshirani1996} as a regression analysis method that performs variable selection. 
The aim (in notation congruent with the deconvolution introduction) is to estimate a sparse vector of linear regression coefficients $s$ in linear regression:
$$ y = Hs + e, $$
where $y$ is the dependent variable, $H$ is a matrix of regressors, and $e$ is an additive i.i.d. Gaussian noise. 
Then the LASSO problem is formulated as
\begin{equation}
\label{eq:lasso} 
\hat s = \arg\min_s \Big[\lambda \|s\|_1 +  \|y - Hs\|^2_2\Big],
\end{equation}
where $\lambda\ge0$ is a tuning parameter and $\|.\|_p$ is a norm in $L_p$-space, i.e. $\|x\|_p =\big(\sum_{i=1}^{T}|x_i|^p\big)^{1/p}$.
This is an instance of convex optimization and quadratic programming.

The Dantzig Selector was proposed in~\cite{Candes2007} as:
\begin{equation}
\label{eq:dantzig} 
\hat s = \arg\min_s \|s\|_1 , \quad \|H^T(y - Hs)\|_{\infty}<\lambda, 
\end{equation}
where $\lambda\ge0$ is a tuning parameter, $\|.\|_{\infty}$ is a norm in $L_{\infty}$-space: $\|x\|_{\infty} = \sup_i |x_i|$. 
Relation of LASSO and Dantzig Selector estimators was discussed e.g. in~\cite{James2009, Bickel2009} and in \cite{Asif2009}.

	\subsection{Selection Criteria}
In the model selection context, we assume that for given data, we obtain a set of candidate models. 
The selection criteria assess the relative quality of the candidate models by finding a compromise between accuracy of the data fit and model parsimony. 
The model with the lowest value of the selection criteria among the tested models is considered the best.

	The \textit{\textbf{Akaike Information Criteria (AIC)}} is defined as
	$$ AIC = -2\ln L + 2k,$$
	where $L$ is the maximised value of the likelihood function of the model, 
	and $k$ is the number of estimated model parameters.

	The \textit{\textbf{Bayes Infromation Criteria (BIC)}} is defined as
	$$ BIC = -2\ln L + k\ln N,$$
	note that it also takes into account the sample size $N$.

The \textit{\textbf{adjusted Akaike Information Criterion (AICc)}} is recommended instead of AIC (see~\cite{Hurvich1989,Burnham2004}) when $k$ is large relative to sample size $N$. It is defined as
$$ AIC_c = AIC + \frac{2k(k+1)}{N-k-1}. $$


For LASSO and Dantzig Selector, AIC and BIC give
\begin{equation}
\lambda^{AIC} = \arg\min_{\lambda} \Big[ N\ln (\lVert y - H \hat s(\lambda)\lVert_2^2) + 2 k(\lambda)\Big], 
\end{equation} 
\begin{equation} 
\lambda^{BIC} = \arg\min_{\lambda} \Big[ N\ln (\lVert y - H \hat s(\lambda)\lVert_2^2) + k(\lambda) \ln N\Big], 
\end{equation} 

where $\hat s(\lambda)$ is the solution of~(\ref{eq:lasso}) and (\ref{eq:dantzig}), that corresponds to $\lambda$, $y$ is the measured signal, $H$ is the matrix of regressors, $k(\lambda)$ is the number of non-zero parameters in $\hat s(\lambda)$ (see~\cite{Zou2007}).
Comparison of using the AIC and BIC to Dantzig Selector in deconvolution tasks in neuroimaging was discussed in~\cite{Gaudes2013}.

\section{MCI}
\label{sec:mic}
		
\subsection{Basic idea: Bayes classifier}

The idea of our proposed approach is to utilize the information available in the observed data and the derived signals (namely the ordinary least square estimate of the brain activity and the family of regularized estimates) within a Bayesian inference tool that separates the true activations from false positives. 
The starting point is a basic estimate of the brain activity $\xi = (\xi_1,\dots,\xi_N)$, $\xi_j \in \mathbb{R}$ (think of e.g. the OLS solution $\hat{s}_{OLS}$). We assume that its non-zero elements correspond to apparent activations due to one of two cases: neuronal response (true activation) or noise (false activation), and apply a naive Bayes classifier to separate the two classes. 

Let us discuss the Bayes classifier in more detail. Let $g : \mathbb{R} \rightarrow \{1,\dots,M\}$ be the function (classifier), which  returns the number of mixture component (class) $g(\xi) \in \{1,\dots,M\}$ for each possible value of the observed characteristics $\xi \in \mathbb{R}$. 
In our case, there are two classes, i.e. $ M=2,\quad g(\xi_j) \in \{1,2\}, $
where $g(\xi_j) = 1$ corresponds to the conclusion '$j$-th element falls to a true activation component', and $g(\xi_j)=2$ corresponds to the conclusion '$j$-th element falls to a false detection component'.
The Bayes classifier is of the form 

\begin{equation}
g^B(\xi_j) = \arg\max_{m=1,\dots,M} \, p_j^m f^m(\xi_j),
\end{equation}

where $j\in\{1,\dots,N\}$ indexes the classified elements of the vector, $\xi_j\in\mathbb{R}$ are the basic (OLS) estimate values, $p_j^m$ are the probabilities for $j$-th activation to fall to the $m$-th class, and $f^m$ denotes the probability densities of the  characteristic $\xi$, provided it falls within the class $m$.
Thus, the classifier $g^B(.)$ for each $j$-th activation compares the posterior probabilities of the element belonging to a neuronal activation or noise.
For details and properties of such a classifier see ~\cite{Rish2001} and~\cite{Hand2001}.

However, true values of the probabilities $p^m$ and the densities $f^m(x)$ are usually unknown a priori. Instead, we may need to obtain some estimates $\hat p^m$ and $\hat f^m(x)$ from data. 
We suggest two approaches to obtain such estimates. First, by using the Gaussian mixture model (GMM), described in~\cite{McLachlan2000}. 
It is based on the assumption of the normal distribution of all mixture components. This approach is well-known, and its main advantage is a developed theory and existence of software tools for the automatic estimation of all the parameters of the model. For our purposes, it is important to estimate both of the needed elements, probability vectors $\hat p^m$ and component densities $\hat f^m(x)$. We test this approach in Section~\ref{sec:simulations}. 

\begin{figure*}
	\begin{center}	
        \includegraphics[width=\textwidth]{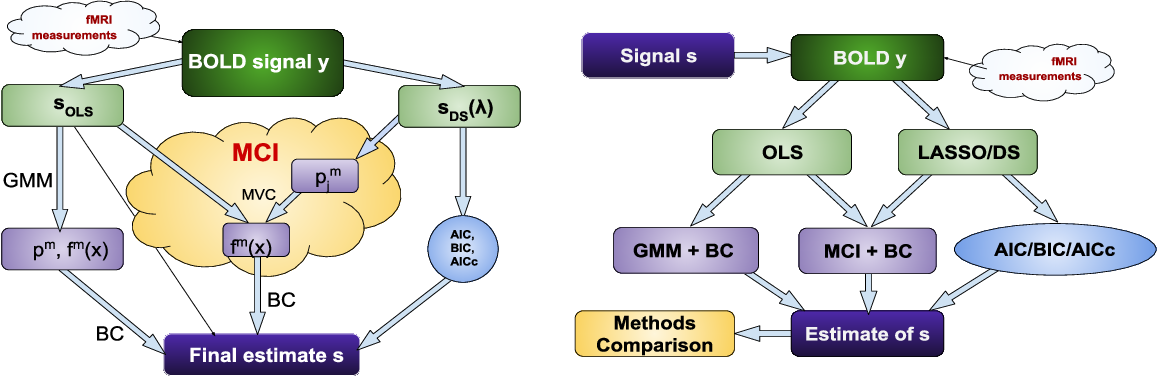}
	\end{center}
	\caption{Left panel: Block-scheme of the compared neural signal estimation approaches. The BOLD signal is first used to estimate the OLS signal estimate $s_{OLS}$ and the family of sparse estimates $s_{DS}(\lambda)$ parametrized by the regularization parameter $\lambda$. Then, either [left branch] the GMM can be applied to the OLS estimate $s_{OLS}$, to provide an estimate for $p^m$, $f^m(x)$ and ultimately a signal estimate, or [right branch] one of the AIC/AICc/BIC selection criteria can be applied to the family of sparse estimates $s_{DS}(\lambda)$ to select a particular sparse estimate, or finally [middle branch], MCI uses $s_{DS}(\lambda)$ to estimate ${p_j}^m$, this is used together with the OLS estimate $s_{OLS}$ to estimate $f^m(x)$, and these are then subjected to a Bayesian Classifier to define the final neural signal estimate. Right panel: scheme of simulation-based comparisons. From the simulated neuronal activity $s$ the BOLD signal $y$ is generated by convolution and adding noise. Then, different sparse methods are applied: LASSO/DS with AIC/AICc/BIC selection criteria, or OLS estimate followed by removing noisy activations using GMM or MCI. These three types of estimates are then compared against the ground truth.
	}
	\label{fig:MCI_block}
\end{figure*}

The second approach is based on utilizing a model with varying concentrations (MVC) for estimating the component densities, and a heuristic procedure based on LASSO/Dantzig Selector algorithms for approximating the probabilities. 
Using this concept enables us, not only to disregard the assumption about the normality of the component distributions, but it also allows for utilizing element-specific prior probabilities -- that are in the present scenario informed by the regularized regressions solutions. 
The detailed description of this concept is provided below and its block-scheme is shown in the left panel of Fig.~\ref{fig:MCI_block}.

\subsection{MVC-based density estimates}

In the classical finite mixture model, the mixing probabilities and thus probability densities of all features $\xi_j$ are supposed to be the same for all the observed objects $j$~\cite{McLachlan2000}:
\begin{equation}
\label{eq:gmm} 
f_j(\xi_j)=f(\xi_j) = \sum_{m=1}^M p^m f^m(\xi_j). \end{equation}

As a generalization of this approach, the model with varying concentrations (MVC) was proposed~\cite{Maiboroda2012}. According to this model, mixing probabilities $p^m_j,\, m\in\{1,\dots,M\},\,j\in\{1,\dots,N\}$, vary for different objects $j$, although still satisfying the condition: $\forall j: 0\leq p^m_{j}\leq1$, $\sum_{m=1}^M p_{j}^m =1.$ Then the distribution for the $j$-th object can be modelled as

\begin{equation}
\label{eq:mvc}
f_j(\xi_j) = \sum_{m=1}^M p^m_j f^m(\xi_j).  
\end{equation}

In MVC, the mixing probabilities $p^m_j$ are supposed to be known, whereas the densities of components $f^m$ are  unknown and estimated from the data using the kernel estimate
\begin{equation}
\hat f^m(x) = \frac{1}{b_mN}\sum_{j=1}^N a^m_j K\Big(\frac{x-\xi_j}{b_m}\Big), 
\end{equation}

where $K(.)$ is a kernel function and $b$ is the bandwidth and $a^m_j$ are the weights, aimed to distinguish the $m$-th component and suppress the influence of other components on it (see~\cite{Maiboroda2012}). 
The parameters are obtained using estimates developed in~\cite{Maiboroda2008} and \cite{Doronin2015}. Classification of components of mixture are discussed in~\cite{Sugakova2006} and in~\cite{Autin2012}.
In particular, let us show the formulas for weights' computation. We denote the inner product of two vectors $a = (a_1,...,a_N)$, and $b=(b_1,...,b_N) $ by $\langle a,b \rangle = \frac{1}{N}\sum_{i=1}^N a_ib_i$ and define the matrix 

\begin{equation*}
\label{eq:weights} 
\Gamma_N = (\langle p^l, p^m\rangle_N)_{l,m=1}^M. 
\end{equation*}

Note that this matrix is assumed to be invertible, with $\gamma_{km}$ being the $km$-minor of $\Gamma_N$. Then the weights are given by:

$$
a^k_{j} = \frac{1}{\det\Gamma_N}\sum_{m=1}^M (-1)^{m+k}\gamma_{km}p^m_{j}. 
$$

For the case of two-components, the weights $a^m_j$ are:

$$
a^1_j = \frac{(1-\langle p^1,1 \rangle) p^1_j+(\langle p^1,p^1 \rangle-\langle p^1,1 \rangle)}{\langle p^1,p^1 \rangle - (\langle p^1,1 \rangle)^2},
$$
$$
a^2_j = \frac{\langle p^1,p^1 \rangle-\langle p^1,1 \rangle p^1_j}{\langle p^1,p^1 \rangle - (\langle p^1,1 \rangle)^2}
,\quad j=1,...,N. 
$$

The kernel width $b_m$ is estimated using the Silverman's rule-of-thumb (see, for example,~\cite{Hardle2006}):
$$
b_m \approx 1.06 \hat \sigma_m N^{-1/5},
$$
where $\hat \sigma_m$ and $\hat \mu_m$ are the estimated standard deviation and mean of the $m$-th component given by $\hat \sigma_m^2 = 1/(N-1)\sum_{j=1}^N a_j^m (\xi_j - \mu_m)^2$, and 
$\hat \mu_m = 1/N\sum_{j=1}^N a_j^m \xi_j$.

\subsection{Probability estimates}
  
As described above, for the classification of signal values $s_j$, we need to set the probability $p^m_j$ that it belongs to the true or false activation class. In this section, we propose a procedure based on the numerical solutions of the optimisation problems with a regularisation parameter, in particular the homotopy algorithms for LASSO~(\ref{eq:lasso}) and Dantzig Selector~(\ref{eq:dantzig}) estimators. 

Let us denote $s(\lambda)$ the solution of~(\ref{eq:lasso}) (or~(\ref{eq:dantzig})) corresponding to the value $\lambda$. The support of $s(\lambda)$ and its sign vector are 
$$\mathcal{B(\lambda)} = \text{supp}\,s(\lambda) = \{j|1\le j \le N, \,s(\lambda)_j\ne 0\}, $$
$$sgn(\lambda) = \{sgn(s(\lambda)_j),\, j=1,\dots,N\}.$$ 
For both mentioned estimators, the particular solution $s(\lambda)$, its support $\mathcal{B}(\lambda)$ and sign vector $sgn(\lambda)$ are specified by the parameter value $\lambda$.
For the LASSO problem, it was shown in~\cite{Efron2004}, that for a given response vector $y$, there is a finite sequence of $\lambda$'s, called transition points (see~\cite{Zou2007}),
$$ \lambda_0 > \lambda_1>\dots>\lambda_K=0, \quad \text{such that:}$$
\begin{itemize}
	\item for all $\lambda>\lambda_0$, solution of the optimization problem (\ref{eq:lasso}) corresponds to the trivial solution, i.e. its support is the empty set  $\mathcal{B}(\lambda) = \emptyset$;
	\item the value $\lambda_K=0$ corresponds to the ordinary least squares solution;
	\item for $\lambda$ values from the interval $I_m := (\lambda_{m+1},\lambda_{m})$, the solutions of (\ref{eq:lasso}) are constant with respect to $\lambda$ in terms of their supports and sign vectors;
	$$ \mathcal{B}_m := \mathcal{B}(\lambda), \quad 
	sgn_m := sgn(\lambda) \quad \forall \lambda\in I_m. $$
\end{itemize}
In other words, there is a sequence $\mathcal{B}_m$, $m = 0,\dots,K$ of subsets of indices $j, j = 1,\dots,N$ that correspond to nonzero coordinates of $s(\lambda)$ for $\forall \lambda \in I_m$.
Thus, for each $j = 1,\dots,N$ there is a set of intervals $I_m$, where the $j$ lies in the corresponding solution support $\mathcal{B}_m$:
$$ \mathcal{I}_j := \cup_{\{m: j\in\mathcal{B}_m\}} I_m  = \cup_{\{m: j\in\mathcal{B}_m\}} (\lambda_{m+1},\lambda_m). $$
Let us denote the union of all intervals $I_m$ as $\mathcal{I} := \cup_{m=0}^K I_m = (0,\lambda_0)$. 
We propose to approximate the probability $\omega_j$ that $j$-th activation comes from the true neuronal signal as

\begin{equation}
\hat  p^1_j = \frac{|\mathcal{I}_j|}{|\mathcal{I}|} = \frac{\sum_{m=0}^{K-1} (\lambda_m - \lambda_{m+1})\mathbb{I}\{j\in \mathcal{B}_m\}}{\lambda_0}. 
\end{equation}

This approach corresponds to the idea, 
that more frequent appearance of non-zero activation estimate in the solution sets means higher probability for this moment to correspond to a true neuronal activation.
Although the probabilistic interpretation of this 'relative occurence in the solution set' is clearly heuristic, it is reasonably motivated as the LASSO operator is assumed to detect true variables among the noise.

Numerically, the set of transition points and corresponding sets of solutions for LASSO can be obtained by the homotopy algorithm LARS, presented in~\cite{Efron2004}.
For Dantzig Selector, there is an available homotopy algorithm  Primal-Dual Pursuit, which was presented in~\cite{Asif2009}. 
Note that the Primal-Dual Pursuit produces a finite sequence of so-called critical values of the parameters, that includes the points of changes in the support or sign vectors of the solutions; without loss of generality, these critical values can be considered as the transition points.

Thus, the probability $p_j^1$ for $j$-th activation to come from true activation, can be approximated based on the numerical solution of the optimisation problem with regularisation parameters. Then, the probability $p_j^2 = 1- p_j^1$, and  the densities $f^1(.), f^2(.)$ can be estimated. Finally, this information is enough to build the classifier and separate the components of vector $\xi$ on true and noisy activations.

\section{Simulation study}
\label{sec:simulations}

To assess the newly proposed methods against the standard methods we first use numerical simulations. We generate a signal $y = s\star h + e$ using HRF $h$, while parametrically varying the the properties of the signal $s$ and noise $e$. 
Note that the Balloon-Windkessel continuous model can be used for generation of even more realistic BOLD signals; we compared it with our data model from Section~\ref{sec:data}, and found out that the correlation between convolutional BOLD and the signal generated using the Balloon-Windkessel model is around 0.99.

After the BOLD signal is generated, LASSO and Dantzig Selector with AIC/AICc/BIC selection criteria are used to estimate the input signal $s$. 
These estimates are compared with two approaches based on Bayes Classifier: GMM (components estimated by~(\ref{eq:gmm})) and the proposed method Mixture Components Inference (MCI, components estimated by~(\ref{eq:mvc})).
To assess the quality of the obtained estimates, several measures are used.  General scheme is shown in the right panel of Fig.~\ref{fig:MCI_block}.

\begin{figure}
	\begin{center}	
        \includegraphics[width=\columnwidth]{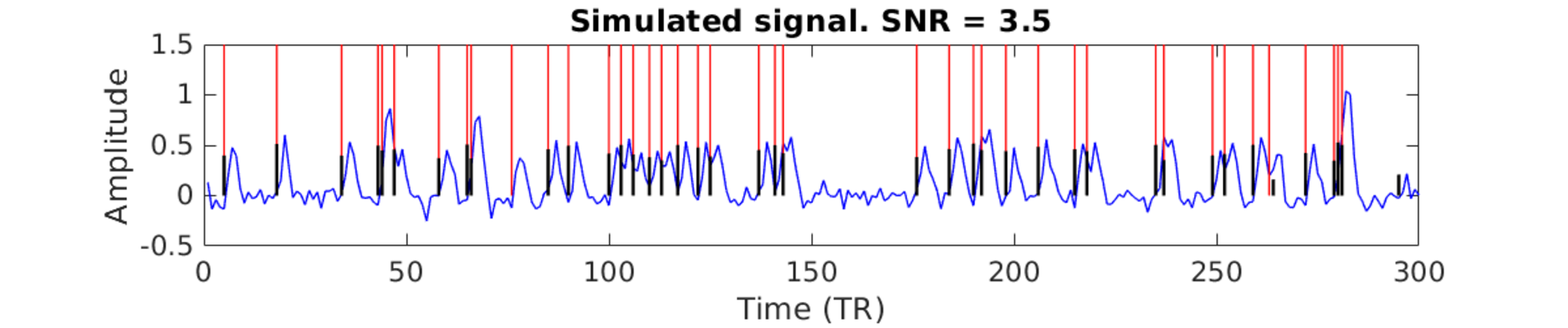}                        
	\end{center}
	\caption{Example of a simulated and estimated signal. The simulated neuronal signal is shown in red, the simulated BOLD signal is shown in blue, the estimated neuronal signal is shown with thick black lines.
	}
	\label{fig:blocksparsesignalDec}
\end{figure}

The HRF kernel is obtained using the function \textit{spm\_hrf(.)} in the SPM toolbox (The Wellcome Dept. of Cognitive Neurology, University College London), setting scan repeat time (RT) to 2.5s. 
To define the level of added noise $e$, we use the Signal-to-Noise Ratio (SNR), $ SNR = \frac{\sigma_{s\star h}}{\sigma_e}$, where $\sigma_{s\star h}$ and $\sigma_e$ are standard deviation of BOLD signal and noise, respectively.
Note that there is a high variability of presented SNR definitions, moreover the values of observed SNR for real data in the literature~\cite{Welvaert2013} likely depend on many aspects of data acquisition, preprocessing and analysis techniques.
For our simulations we used values of SNR in the range $[0.8,3.5]$. This rangewas chosen based on SNR estimated for the datasets presented in the current paper as well as in~\cite{Pidnebesna2018a}.
The input signal $s$ was simulated as a vector of the length $N$, that consists of $K$ non-zero activations (value 1) and $N-K$ zero values. Positions of non-zero elements are chosen uniformly in the range $[1,\dots,N]$.
Fig.~\ref{fig:blocksparsesignalDec} shows an example of a simulated signal and the obtained estimate.
The value of the selection criteria depends on the length of the signal $N$ and number of non-zero elements $K$. Thus, we explore the estimates for several different noise levels and multiple numbers of true non-zero coefficients.
To assess the model quality we use the Jaccard index, sensitivity, and specificity.

\subsection{Numerical results}

In this section, we present the numerical results of the simulation study.
First, our simulations show that DS/LASSO with AIC/BIC tends to mark almost all times  as activated, in other words, extremely over-fits the data. This observation is illustrated by the figures in the Supplementary material, Section D. Thus, the rest of this section presents only the results for DS AICc, MCI DS, and Bayes classifier based on GMM components (BC GMM).

\begin{figure*}
	\begin{center}	
        \includegraphics[width=0.95\textwidth]{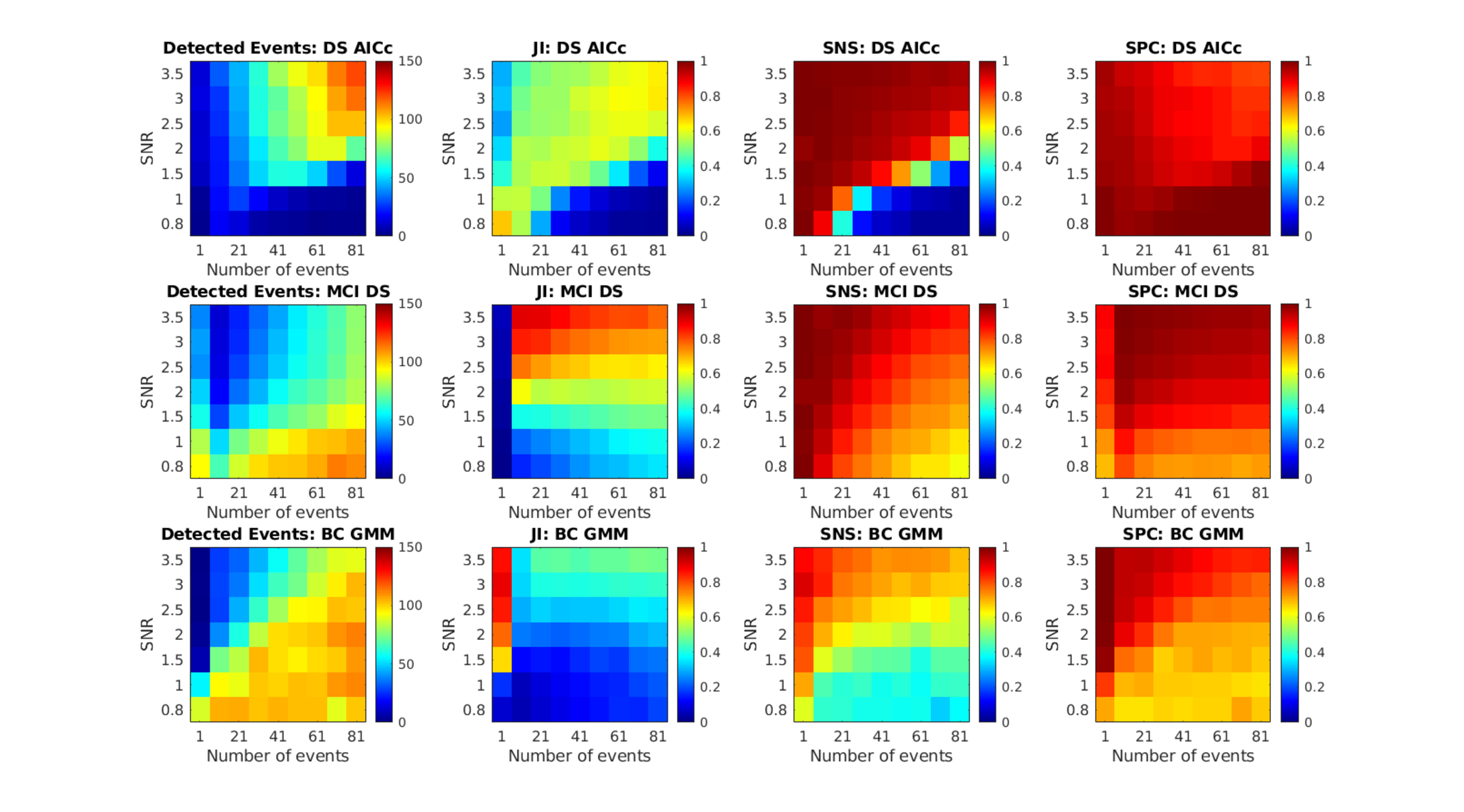}                
	\end{center}
	\caption{Comparison of the performance of the deconvolution methods using simulated data in terms of the number of detected events, Jaccard index (JI), sensitivity (SNS), and specificity (SPC). 
	Time series length is $N=300$, the number of non-zero peaks is shown on the $x$-axis, the noise level on the $y$-axis. Estimation methods are shown in different rows: DS AICc, MCI DS and BC GMM. }
	\label{fig:SNS_SPC_sparse}
\end{figure*}

Fig.~\ref{fig:SNS_SPC_sparse} shows the averaged results for 100 simulations for varying noise levels and activation density. 
Clearly, BC GMM works well only for very sparse signals, and moreover tends to over-estimate the signal for lower SNR.
Note that using a uniform distribution between the noise and activation class (i.e. in the sense $p^1 = p^2 = 0.5$) leads to even worse performance. In particular, GMM with optimal choice of $p^m$ outperforms the $50/50$ version for sparse signals, and the results are comparable or even slightly better in the simulation settings where the actual probability of peaks is close to $0.5$.

The other two methods show generally better results. DS AICc tends to slightly over-estimate the number of detected events for higher SNR and less sparse signals, whereas it severely underestimates for low SNR. Otherwise, the method performs relatively well and gives stable results for the presented parameters. The Jaccard index is typically about 0.6, which is much better than the results of BIC or uncorrected AIC.

The MCI DS shows generally the best results as illustrated by the almost uniformly dominant Jaccard index, with balanced performance in terms of sensitivity and specificity. Indeed it does fail only in the case of extremely sparse signal (simulated by the case of a single activation from 300 points) -- this happens due to inability to estimate the distribution of true activations using only one point.

For an approximate measure of performance differentiation between the methods one can state the minimal signal to noise ratio, for which the Jaccard index reaches 0.8. Note that the MCI method reaches this performance at about SNR=3 across a range of true sparsities of the activations, the DS never reaches this performance, while the GMM shows this performance only in the case of a single true event. For a richer picture, however, one needs to explore the dependence of the performance on the sparsity of the true events, as well as the required tradeoff between sensitivity and specificity, so the results can’t be easily summarized in one measure.
 

\section{Real data examples}
\label{sec:experiments}

For validation of the proposed deconvolution approach on real data, we use two different datasets with distinct experimental designs and data acquisition parameters. In particular, we use an internal dataset of subjects performing a memory task and a publicly available dataset originating from the Human Connectome Project involving motor task fMRI data.

\subsection{Working memory task}
The first used dataset has been collected in the National Institute of Mental Health (Czech Republic). Only the measured fMRI BOLD signals are used for the deconvolution procedure, and information about subjects' responses are used for quality assessment of MCI in comparison with LASSO/DS. 

\subsubsection{Working memory task: Subjects}
The real data example includes the data of 56 subjects (24 males and 32 females, age = $33.66 \pm 10.97$) that performed fMRI recording with n-back task as a part of a complex study with repeated psychiatric, neuroanatomical and neuropsychological evaluation. All subjects have been recruited for the study based on the following exclusion criteria: history of mental disorder, history of neurological disorder, presence of any artificial objects that would interfere with the magnetic resonance imaging. The study was conducted in accordance with the Declaration of Helsinki and approved by the local Ethics Committee of the National Institute of Mental Health (Czech republic). All participants were informed about the purpose of the study, the experimental procedures, as well as the fact that they could withdraw from the study at any time, and provided written informed consent prior to their participation. The data set of 89 MRI sessions in total was analyzed, including data from repeated sessions (6 months apart) in 25 subjects (2 visits for 17 subjects and 3 visits for 8 subjects).

\subsubsection{Working memory task: Experimental design}
The n-back task is a continuous performance task commonly used for assessment of working memory performance.
The subject is presented with a sequence of stimuli, and required to indicate when the current stimulus matches the one from n steps earlier in the sequence. 
The individual button press timings are used as a proxy for true activations of the motor cortex for individual analysis. As all subjects had the same experimental design, we also carry out a group level analysis, setting as group ground-truth as button presses present in at least 80\% of subjects (71 of 89) and averaging the measured BOLD signal in order to decrease noise. 
On average the subjects responded by button press 44.63 times (SD = 6.18, min = 27, max = 58). Most of the responses were correct, on average 5 of the responses were incorrect.

\subsubsection{Working memory task: fMRI acquisition and preprocessing}
Brain images were obtained using Siemens Prisma 3T MR machine. Functional T2*-weighted images with BOLD contrast were acquired with voxel size 3x3x3 mm, slice dimensions 64x64 voxels, 32 axial slices, repetition time 2000 ms, echo time 30 ms, flip angle $70^{\circ}$. Then, the standard fMRI preprocessing was done using the SPM8 toolbox with spatial realignment, slice-time correction, normalization to the standard anatomical space, and spatial smoothing. The primary motor cortex time series was extracted by independent component analysis using the GIFT toolbox.

\subsubsection{Working memory task: Results}

\begin{figure}
	\begin{center}	
        \includegraphics[width=\columnwidth]{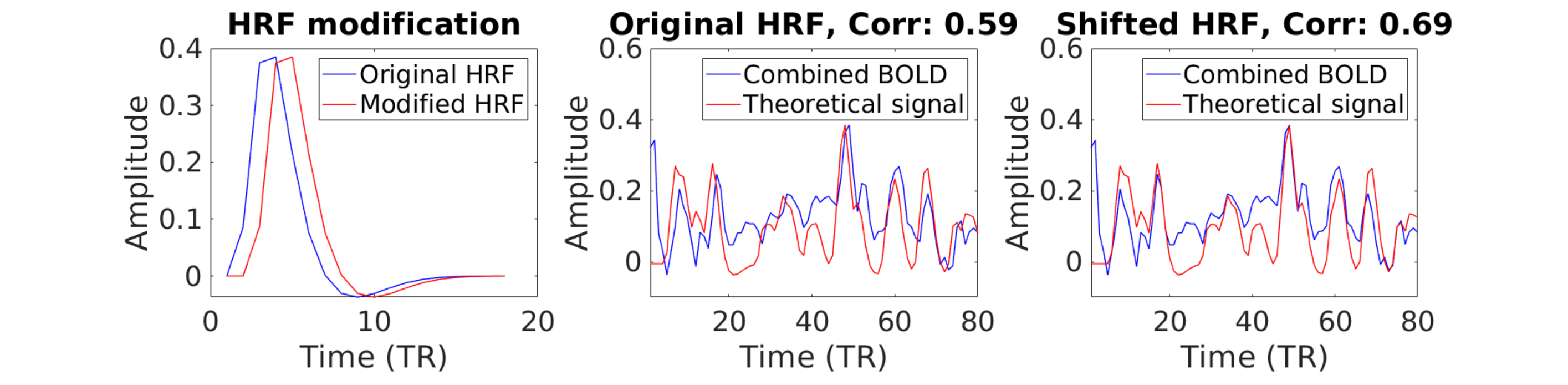}        
	\end{center}
	\caption{Selection of appropriate hemodynamic response function (HRF) using a forward model on real data from the working memory task. Left: the default and shifted HRF. Comparison of the averaged measured BOLD signal with the convolution of the true activations vector and the original HRF (middle) and shifted HRF (right). Note that the visually negligible effect of shifting the model by one TR has substantial effect on the agreement (correlation) between measured and theoretical BOLD signal.}
	\label{fig:Nback_correlation}
\end{figure}

To test that we use a suitable HRF kernel, we have first applied the forward model using the standard HRF function available in the SPM software, as well as its versions shifted by 1 TR backward and forward. 
By comparing the measured signal with the theoretical one, we found out that the data is in line with the HRF delayed by 1 TR (shifted HRF), rather than the default HRF, see  Fig.~\ref{fig:Nback_correlation}.
We further compared our methods MCI DS and MCI LASSO with the standard approaches DS and LASSO with the selection criteria AIC, AICc, BIC.

%

%

\begin{figure}
	\begin{center}	
        \includegraphics[width=\columnwidth]{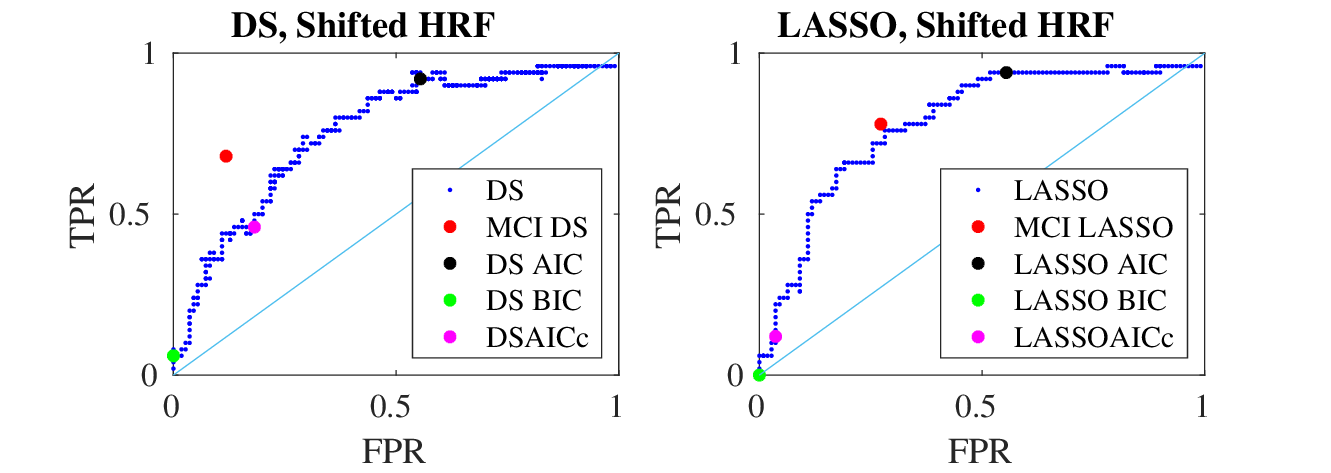}                
	\end{center}
	\caption{
	ROC curve of the Dantzig Selector and LASSO solutions for the group-level real data signal from the working memory task. The blue line of points corresponds to the set of solutions obtained by LASSO/DS as function of its one free parameter. The other colored points specify the particular solution selected by the selection procedures (in case of AIC, AICc, BIC, DS), or the newly proposed method (MCI).}
	\label{fig:Nback_ROC_combined}
\end{figure}

Numerical results for the combined (group) signal are shown in Table~\ref{tab:SCcomparison_HRF}, and the corresponding receiver operating characteristic (ROC) curve is shown in Fig.~\ref{fig:Nback_ROC_combined}. 
For DS BIC, the estimate is almost a trivial vector. It contains only true activations, however only a fraction of them. Conversely, the DS AIC-estimate includes more true activations (high sensitivity), but also contains many false detections. AICc suffers from mediocre sensitivity. 
MCI DS then shows the highest values of all three quality characteristics. LASSO shows a similar picture. 

\begin{figure}
	\begin{center}	
        \includegraphics[width=\columnwidth]{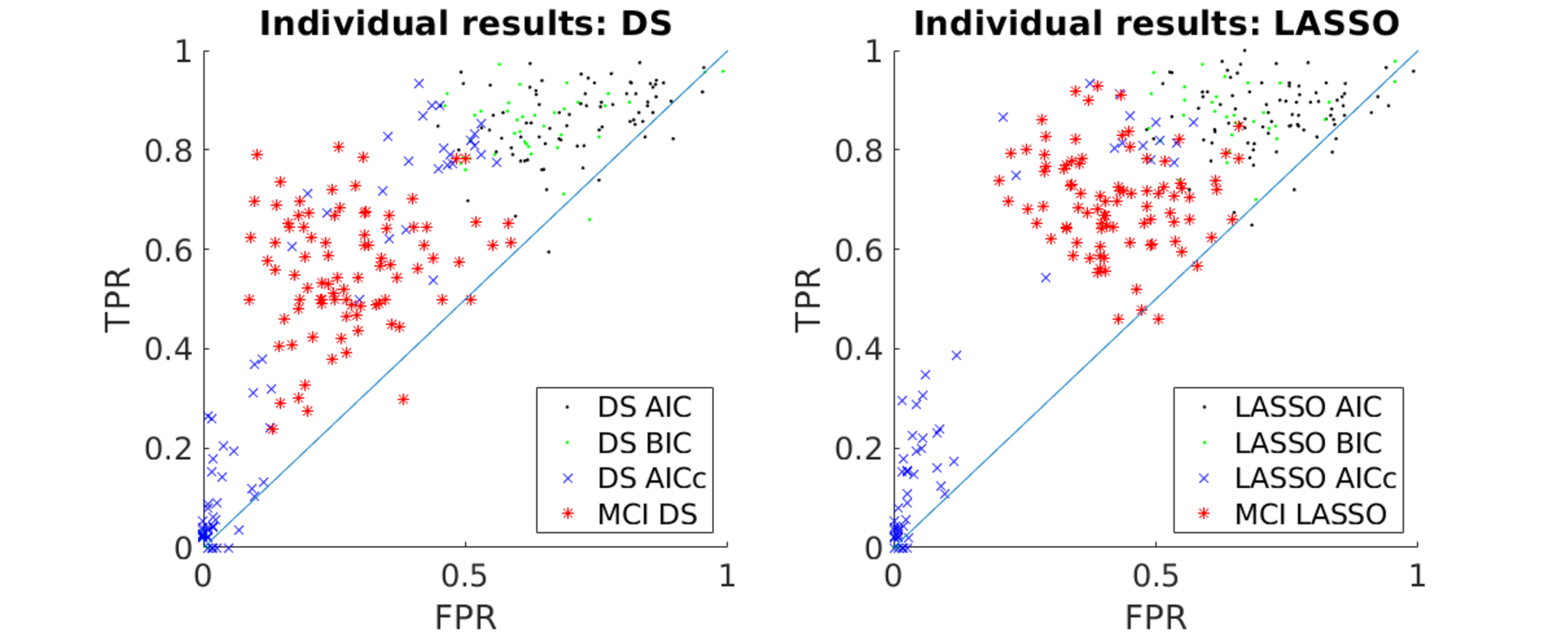}          
	\end{center}
	\caption{Comparison of deconvolution methods performance on individual-subject level - results from the working memory task. Left: results for the Dantzig Selector, right: results for LASSO.} 
	\label{fig:Nback_AllSbj_ROC_1hrf}
\end{figure}

Fig.~\ref{fig:Nback_AllSbj_ROC_1hrf} shows the quality of estimates of each subject's time series. The black and green points that mark DS AIC and DS BIC methods form similarly placed clouds and demonstrate that almost all estimates contain too many positive detections. The blue crosses mark the DS AICc and show a lot of estimates with too few true positive detections. The rest of the blue crosses forms a cloud that has an intersection with all other methods. Red color is used for MCI DS estimates, and this cloud lies away from the DS AIC and BIC and demonstrates the most appropriate estimates in terms of true and false positive rates. See table~\ref{tab:subjectLevel} for summary results.

\begin{table}
\caption{Summary group-level performance: working memory task.}
	\label{tab:SCcomparison_HRF}
	\begin{center}									
		\begin{tabular}{|l|ccc|}
			\hline							
			&	JI	&	SNS	&	SPC	\\
			\hline							
			DS BIC	&	0.07	&	0.07	&	1.00\\
			DS AIC	&	0.35	&	0.87	&	0.41	\\
			DS AICc	&	0.31	&	0.46	&	0.81	\\
			\textbf{MCI DS}	&	\textbf{0.51}	&	\textbf{0.61}	&	\textbf{0.92}	\\
			\hline							
			LASSO BIC	&	0.00	&	0.00	&	1.00	\\
			LASSO AIC	&	0.42	&	0.94	&	0.45	\\
			LASSO AICc	&	0.11	&	0.12	&	0.96	\\
			\textbf{MCI LASSO}	&	\textbf{0.49}	&	\textbf{0.78}	&	\textbf{0.73}	\\
			\hline							
		\end{tabular}								
	\end{center}									
JI = Jaccard index, SNS = sensitivity, SPC = specificity. The novel MCI method results in bold.							
						
\end{table}

\begin{table}
\caption{Summary subject-level performance: working memory task.}	
\label{tab:subjectLevel}
	\begin{center}									
		\begin{tabular}{|l|ccc|}
			\hline							
			&	JI	&	SNS	&	SPC	\\
			\hline							
			DS BIC	&	0.14(0.12)	&	0.37(0.38)	&	0.72(0.38) \\
            DS AIC	&	0.30(0.04)	&	0.86(0.12)	&	0.30(0.11)	\\			DS AICc	&	0.14(0.11)	&	0.26(0.25)	&	0.86(0.24)	\\
			\textbf{MCI DS}	&\textbf{0.33}\textbf{(0.05)}&\textbf{0.56}\textbf{(0.12)}&\textbf{0.72}\textbf{(0.10)}	\\
			\hline							
			LASSO BIC	&	0.12(0.16)	&	0.32(0.42)	&	0.76(0.32) \\
            LASSO AIC	&	0.30(0.05)	&	0.87(0.07)	&	0.28(0.11)	\\		LASSO AICc	&	0.12(0.15)	&	0.19(0.30)	&	0.91(0.16)	\\
			\textbf{MCI LASSO}	&\textbf{0.34}\textbf{(0.08)}&\textbf{0.70}\textbf{(0.10)}&\textbf{0.58}\textbf{(0.11)}	\\			
			\hline							
		\end{tabular}								
	\end{center}									
JI = Jaccard index, SNS = sensitivity, SPC = specificity. Mean (std) across subjects is shown for each method. The novel MCI method results in bold.							
								
\end{table}	

\subsection{Motor task}

As a second dataset for our analysis, we use the task-related data from the Human Connectome Project. We choose this dataset as it is publicly available, and allows testing the role of specific artifacts due to the acquired physiological data.

\subsubsection{Motor task: Subjects}
We analyze preprocessed task-related fMRI data of 62 subjects from the Human Connectome Project~\cite{Vanessen2013}, \url{https://humanconnectome.org/study/hcp-young-adult}. 
One of the goals of this analysis is to study the influence of the physiological processes (breathing and heartbeat) on the deconvolution quality. Thus, from the 100 downloaded subjects (Motor task), we choose only those for whom corresponding physio-data were available.
The analysis is performed on the averaged signals from primary motor cortex (Broadman area 4) for both hemispheres (the LR-modality). The original sample rate of the data is $0.72s$, however, we subsample the data to $TR = 2.16s$ for comparability with our previous dataset. 

\subsubsection{Motor task: Experimental design}
For our analysis, we choose the motor task. 
Volunteers were asked to tap their left or right fingers, or squeeze their left or right toes, or move their tongue to map motor areas. Each block included 10 movements (12 seconds). In total there were 13 blocks, with 4 of hand movements (2 right and 2 left), 4 of foot movements (2 right and 2 left), and 2 of tongue movements. In addition, there were 3 15-second fixation blocks. 
Detailed description of the experimental paradigm can be found in \cite{Barch2013}. 

\subsubsection{Motor task: Physiological data}

It is widely known that physiological fluctuations resulting from the heartbeat and respiration are a substantial source of noise in fMRI, as it perturbs BOLD by fluctuations of non-neuronal origin~\cite{Kruger2001,Hutton2011,Triantafyllou2005,Liu2016}. This causes challenges in the analysis, in particular, false-positive results may emerge if non-neuronal physiological fluctuations correlate with neuronal activation.

A commonly used physiological noise correction techniques for fMRI data include model-based RETROspective Image CORrection (RETROICOR), using respiratory volume per time and heart rate variability responses (RVT/HRV)~\cite{Glover2000}. 
In our analysis, we use the mentioned techniques implemented in the PhysIO toolbox, a part of the TNU "TAPAS" software suite~\cite{Kasper2017}. PhysIO is used for obtaining the matrix of physiological regressors, which is later used for cleaning the BOLD signal from breath/heartbeat influence for every subject.

\subsubsection{Motor task: Analysis and Results}

For every subject, we extract two time-series corresponding to the primary motor cortex (left and right hemisphere) and a matrix of regressors that consists of physiological regressors obtained by PhysIO, and the motion regressors as provided for HCP data. The mentioned BOLD signal, obtained after standard HCP preprocessing, we refer to as the \emph{original} signal; also, we run the linear regression with described physiological regressors, to get the \emph{cleaned} BOLD signal. 
In this section we present only the results obtained by Dantzig Selector on the right hemisphere time series. 

\begin{figure}
	\begin{center}	
        \includegraphics[width=\columnwidth]{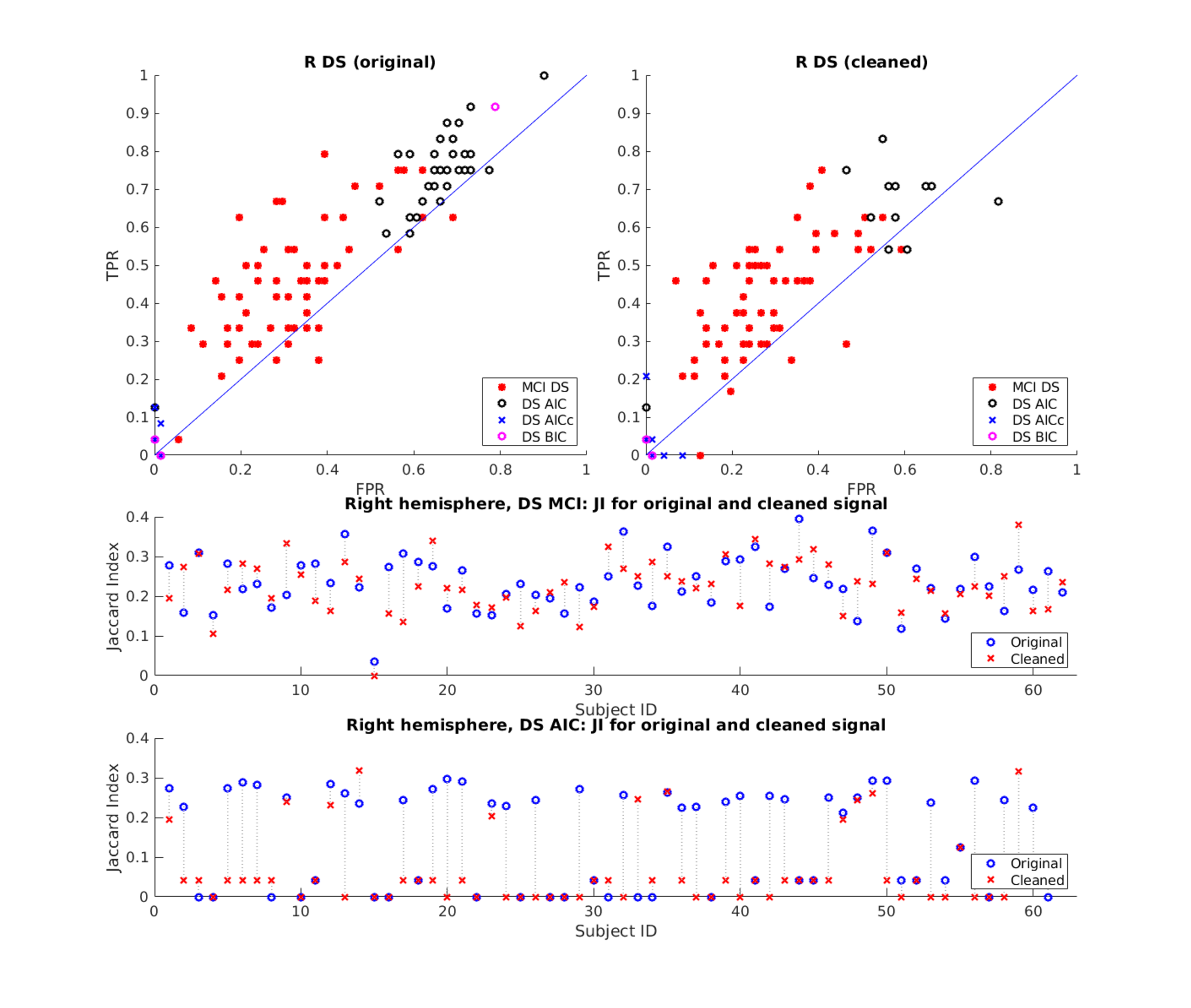} 
	\end{center}
	\caption{Deconvolution results for the motor task data (right hemisphere) from the HCP dataset, focusing on the influence of physiological artifacts on BOLD. Top panel: results of deconvolution of original (left) and cleaned (right) signal. Four different methods of deconvolution are marked in different colors. Middle and bottom panels: Jaccard Index for original and cleaned signal, DS MCI, and DS AIC respectively.}
	\label{fig:Motor_JI}
\end{figure}

We use blocks of movements as known stimuli. Note that we do not distinguish activations caused by hand and foot or tongue. 
We deconvolve both types of signal, original and cleaned, for both hemispheres with all discussed methods; subject-level results can be found in Fig.~\ref{fig:Motor_JI}.
Similarly to the working memory task, DS BIC and AICc show robust underestimate (basically, both of these criteria choose almost trivial solution), AIC tends either to choose a trivial solution or overestimate it by choosing almost all the peaks as activation. MCI DS solutions, marked by red stars, perform relatively reasonable in terms of both, TPR and FPR. 

As another check of how much the cleaning of the signal from physiological and motion artifacts influences deconvolution results, we compare distribution of JI before and after cleaning. Paired $t$-test is used for comparison, showing the significant difference only for DS AIC right hemisphere ($p<0.001$);
similar results are obtained for the LASSO approach instead of DS.


Additionally to the subject-level analysis, we do also group analysis. On a group level, we deconvolve averaged original signal and averaged cleaned signal. 
While the standard criteria of AIC, BIC and AICc fail to choose the appropriate solution, the MCI approach shows a relatively good Jaccard Index ($0.49$ for original signal and $0.48$ for cleaned data). The method clearly does not work perfectly, but performs much better than standard ones. 
Our results, obtained on the subject- and group-level, suggest that for reasonably good quality data with appropriate preprocessing the results are not dominantly affected by false-positives activations due to physiological artifacts.

Finally, we compare JI for MCI DS with SNR for original and cleaned signals. We see a highly significant correlation of the Jaccard Index with SNR for both, original and cleaned versions of BOLD ($r = 0.41$, $p = 0.0014$ and $r = 0.53$, $p = 0.0025$  respectively). This is an additional confirmation of earlier stated conclusion: deconvolution works better on good signals. In general, results obtained on the HCP dataset are in line with the previous ones from the working memory dataset, despite experimental design does not correspond to exactly sparse neuronal activations, but rather continuous blocks of activity. The proposed MCI approach shows more relevant deconvolution results, whereas the typical behavior of classical selection criteria AIC/BIC/AICc is either to overestimate or underestimate the signal for Dantzig Selector/LASSO.

\section{Discussion and conclusions}
\label{sec:discussion}

We have introduced a novel method for recovering a sparse true signal from smoothed and noisy observations, exemplified by estimation of neuronal activity from BOLD signal. The main idea of the approach is to overcome the challenges of regularization combined with standard model selection criteria, which is achieved by aggregating the information contained in the solutions, instead of selecting a single one favoured by the criterion.  Practically this entails applying additional operations on the result of standard regularization-based deconvolution methods, such as LASSO and Dantzig Selector, and thus builds upon previous achievements in the field. Notably, albeit the proposed sparse regression method is introduced in the context of a particular application (fMRI BOLD deconvolution), it constitutes a novel general approach that may be applicable in a variety of other contexts. 

The advantages of this method over the application of the Bayes information criterion and Akaike's information criterion (and its improved variant) were demonstrated both in numerical simulations as well as in experimental data, where it provided more accurate estimates of timing of motor responses occurring during simple motor or more complex cognitive tasks. This provides a proof of principle that the method would improve the detection of timing of neuronal activation in other experimental situations, where the brain activity timing is not directly controlled by the experimental procedure. Notably, the proposed MCI method is comparable with a standard application of the standard regularization approaches in terms of computational demands. In particular, on top of the computationally relatively demanding regularization step, it only adds a substantially less demanding follow-up procedure leading to the final results. 

The general Mixture Components Inference method was exemplified here by application to a particular neuroimaging task -- the estimation of the underlying neuronal activity from fMRI BOLD signal. The BOLD signal character poses several main challenges to such deconvolution task: its low sampling rate compared to the scale of neuronal activity, its indirect and postponed dependence on neuronal signal (hemodynamic response), and complex nature of sources driving it. Low sampling rate is in principle a major limitation compared to for instance electrophysiological measurements. However due to the sluggishness of the hemodynamic response, the BOLD signal anyway corresponds to a smoothed version of the otherwise fast-scale neuronal activations. BOLD signal thus corresponds to relatively long-scale weighted temporal averages (compared to the neuronal firing timescales), justifying thus using relatively low temporal sampling in both the measurements and models. It is customary to impose for convenience the same temporal discretization for the estimated neural activity as for the BOLD measurements, and multiple true activations within a single TR would thus be, even in the best scenario of perfect estimate, be interpreted as a single joint activation in the corresponding time point. Of course, in principle the model could be extended to include finer temporal sampling of the estimated neural activity, however, given the smoothness of the signal and unknown precise hemodynamic response, it might remain challenging to provide
reliable estimates with a sub-TR temporal precision.

Future work may involve combining the presented approach with some other directions recently taken up in the field, namely spatial regularization,  change sparsity, HRF estimation and further processing. In more detail, the possible extensions and integrations include: the joint estimation of the HRF and neuronal activity, such as in the fused LASSO and Tikhonov regularization of the HRF~\cite{Aggarwal2015}, spatial structure modelling including regularization as in the work of~\cite{Karahanoglu2013}, which was further extended in~\cite{Farouj2017} to remove the need for a priori brain parcelation, or alternative penalty schemes in regularization such as combining total variation and negativity penalties as in~\cite{Hernandez-Garcia2011}. Of interest could be the extension to application of such deconvolution using more complex data models, such as the bilinear model including neuronal activity autocorrelation, stimulation and modulation terms~\cite{Penny2005, Makni2008}, potentially in the context of network analysis. Last but not least, the model could be extended to accommodate variability of hemodynamic responses functions, either in relation to space or type of cognitive function; albeit with obvious technical challenges in case of large collinearity the brain activities or the response functions.  

As the outlined approach relies on being combined with an initial method providing a parameterized family of solutions (in this paper, we used the LASSO and Danzig Selector to provide examples of such families), we leave open the question of how it relates and could synergize with other families of approaches such as using the cubature Kalman filtering~\cite{Havlicek2010}, activelets (wavelets designed using prior knowledge on the hemodynamic response function form~\cite{Khalidov2011}), spatiotemporal HRF deconvolution~\cite{Aquino2014}, multivariate semi-blind deconvolution~\cite{Cherkaoui2021}, improved sparse  paradigm  free  mapping with no selection of the regularization parameter \cite{Urunuela2020} or nonparametric hemodynamic deconvolution
of fMRI using homomorphic filtering~\cite{Sreenivasan2015}. We can generally speculate that the key idea of combining model outputs using e.g. Bayesian approaches instead of crisp model selection within a constrained parametric family may be applicable to other schemes beyond classical regularization schemes such as LASSO, as long as these involve selection among parametric model families.

The obtained information on local neural activations can be of interest on its own such as in the application to study cortical response to unpredictable mental events as outlined in~\cite{Gaudes2011}, or possibly used for obtaining meta-analytic interpretation of the brain activation patterns~\cite{Tan2017}. Importantly, it has been proposed that functional connectivity in MRI is driven by spontaneous BOLD events detectable by deconvolution methods~\cite{Allan2015}; and the use of deconvolution as an intermediate step to estimate effective connectivity has been advocated~\cite{Wu2013}. Similarly, the dynamics of deconvolution-based brain activation patterns can be further explicitly modelled to fit sparse coupled hidden Markov models~\cite{Bolton2018} characterizing the large-scale brain dynamics.

Moreover, deconvolution techniques are used in a range of specific applications in neuroimaging. In~\cite{Veloz2019}, authors are using the deconvolution for ROI identification. Deconvolution for multi-echo functional MRI is discussed in~\cite{Caballero-Gaudes2019}, and the authors claim that the technique is very promising in understanding the dynamic nature of brain activity in resting-state and paradigms with unknown timing of the BOLD events. It is also used for computation of activity maps for getting the dynamical functional connectivity in spontaneous flow of thought/resting-state studies~\cite{Gonzalez-Castillo2019}. Authors of~\cite{Deco2017a} mention that even though they use thresholding of the BOLD signal, there is a place for the application of more sophisticated methods as deconvolution. Including deconvolution is also mentioned as possible future work for filtering of the physiological noise~\cite{Sarkka2012}. Improving the signal-to-noise ratio of the data with the deconvolution is pointed out as an option in~\cite{Fallani2014}. Deconvolution has also been used for obtaining the effective connectivity matrix in an autism study~\cite{Deshpande2013}.

Last but not least, the development of deconvolution approaches is relevant not only for processing of most common fMRI BOLD signal but also in the context of functional near-infrared spectroscopy~\cite{Santosa2019, Seghouane2019} and quantitative susceptibility mapping~\cite{Costagli2019}. In general, the neuroimaging field just exemplifies the fruitful interaction between advanced engineering attempts to tackle experimental and technical challenges specific to each measurement modality (or even particular type of experiment) and the development of novel algorithmic and mathematical approaches applicable across disciplines.

\section{Declaration of Competing Interest}
The authors herein declare no conflicts of interest concerning publication of this article.

\section{Acknowledgement}
We thank David Tome\v{c}ek for preparing example fMRI data and Jessica Barilone for helping with language and style editing. This work was supported by the Czech Science Foundation project No. 21-32608S and by Ministry of Health Czech Republic DRO 2021 (``National Institute of Mental Health – NIMH, IN: 00023752'').  The data that support the findings of this study are available from the corresponding author upon reasonable request.

\bibliographystyle{ieeetr}

\end{document}